*Article*

# A Multi-View Framework to Detect Redundant Activity Labels for More Representative Event Logs in Process Mining

Qifan Chen [1]* , Yang Lu [1] , Charmaine S. Tam [2]  and Simon K. Poon [1]

1   School of Computer Science, The University of Sydney, NSW 2006, Australia; yalu8986@uni.sydney.edu.au (Y.L); simon.poon@sydney.edu.au (S.K.P)
2   Centre for Translational Data Science and Northern Clinical School, The University of Sydney, NSW 2006, Australia;  charmaine.tam@sydney.edu.au  (C.S.T)
*   Correspondence: qifan.chen@sydney.edu.au

**Abstract:** Process mining aims to gain knowledge of business processes via the discovery of process models from event logs generated by information systems. The insights revealed from process mining heavily rely on the quality of the event logs. Activities extracted from different data sources or the free-text nature within the same system may lead to inconsistent labels. Such inconsistency would then lead to redundancy in activity labels, which refer to labels that have different syntax but share the same behaviours. Redundant activity labels can introduce unnecessary complexities to the event logs. The identifications of these labels from data-driven process discovery are difficult and rely heavily on human intervention. Neither existing process discovery algorithms nor event data preprocessing techniques can solve such redundancy efficiently. In this paper, we propose a multi-view approach to automatically detect redundant activity labels using not only context-aware features such as control–flow relations and attribute values but also semantic features from the event logs. Our evaluation of several publicly available datasets and a real-life case study demonstrate that our approach can efficiently detect redundant activity labels even with low-occurrence frequencies. The proposed approach can add value to the preprocessing step to generate more representative event logs.

## 1. Introduction

Process mining combines traditional model-based process analysis and data-centric mining techniques [1]. It is a technology known to be useful for understanding business processes and constructing process models using event logs captured in information systems [2]. Process mining includes process discovery, conformance checking, and enhancement [3]. Among them, process discovery is a paramount task that aims at automatically discovering process models from structured event logs to analyse and improve the internal business processes [4]. Process mining has shown promising potential in many aspects, such as discovering significant insights and improving process performances [5]. A typical event log refers to a collection of events, each with a timestamp that records the executed time. An event represents a unique execution of an activity, which is a well-defined step in the process, such as "doctor appointment". Cases group these events, also called process instances. For example, a case could be a patient who follows a treatment process in a hospital.

In recent years, with an effort being made to discover accurate and comprehensible process models from structured and clean event logs, many advanced discovery algorithms have been proposed, such as the Heuristic Miner [6] and Split Miner [7]. However, like other data mining technologies, the quality of input data (event logs) has a great impact on the resulting model [2]. Moreover, real-life event logs can suffer various data quality issues and lead to "spaghetti-like" business models, which refer to models that have numerous activities and intricate relations. Such "spaghetti-like" models are too complex to comprehend easily even for domain experts [8]. The process mining manifesto [9] has emphasised the importance of event log quality in process discovery. The first guideline for process mining is to treat event data as first-class citizens. Suriadi et al. [10] summarised 11 event log imperfection patterns and pointed out that data quality on activity labels in event logs is unique in process mining research, as the quality can be affected by integrating data sources or discrepancies of labelling in the same information system. One particular imperfection pattern is redundant activity labels, which refers to labels that have different syntax but share the same behaviours [11]. Two labels are recognised as sharing the same behaviours not only if they play identical roles in process models but also if they represent the same activity in reality. One contributing factor to such redundancy is data integration from separate systems, e.g., Electronic Medical Records (EMR) from different hospitals, because multiple systems use different labels for the same activity. The other is the free-text input or human error in providing an initial suggestion in the same system [11]. Redundant activity labels are commonly observed in real-life event logs. For instance, nearly 30% of activity labels in observation tests are considered redundant in the publicly available MIMIC-III database [12]. Event logs with such redundant activity labels can suffer a significant quality loss in the discovered process model, because most existing process discovery algorithms assume that every activity label in the event log is meaningful and unique. Several methods are proposed in the field of process mining to detect redundant activity labels in event logs [11,13,14]. However, none of these methods can accurately detect redundant activity labels in event logs without domain knowledge, especially when the redundant activity labels are less frequent and contain numerical values as attributes.

In this paper, we propose a multi-view framework that efficiently incorporates the control–flow relations, attribute values, and label semantic information of each activity label to detect redundancy in event logs. A consensus is then guided by a majority voting mechanism to integrate results produced from multiple views.

### 1.1. Motivating Example

To demonstrate how redundant activity labels can introduce unnecessary complications to discovered process models and the motivations behind our proposed approach clearly, we describe a simple patient treatment process as an example. Assuming there are eight activities in the process, that is, (A) registration, (B) visiting the doctor, (C) performing colonoscopy, (D) performing a laboratory test, (E) performing an MRI, (F) performing surgery, (G) paying the bill, and (H) discharging the patients. Hypothetically, the clean event log contains eight traces denoted by $L_1$ ={ABDEFGH, ABCEFGH, ABCDEFGH, ABFGH, ACEFH, BADFGH, BACDEGH, ABCFGH}. The goal of automated process discovery algorithms is to construct a process model that can accurately describe the process behaviours [15]. For instance, if we apply the popular algorithm used by Disco, which is a tool widely used to generate visual and actionable insights for process mining [16], we can obtain the process model as shown in Figure 1. The numbers in each box indicate the case coverage in Disco. It is easy to interpret the process model: patients usually register first and then visit the doctor. After that, they may be asked to perform a colonoscopy, laboratory tests, or an MRI. A surgery follows depending on the patient's situation. Finally, they pay the bills and are discharged from the hospital. However, redundant activity labels may exist if the event log contains merged data from sources that do not share a common schema [10]. Therefore, the same real-world activity is recorded with different labels in each source. Suppose there are two activity labels, (B1) "DrSeen" and (B2) "Medical Assign", which both represent activity (B) in the event log $L_2$ [10]. Moreover, activity labels (H1) "Release C" and (H2) "Release D" represent the same method by which patients are discharged [17]. The event log is denoted by $L_2$ ={ABDEFGH, AB1CEFGH1, AB2CDEFGH2, ABFGH, ACEFH, B2ADFGH2, B1ACDEGH1, ABCFGH}. Consequently, we obtained another process model based on this redundant event log, which is shown in Figure 2. Comparing the two process models, it can be seen that such redundancy brings unnecessary relations (red arcs) and unwanted activity labels (activity B1, B2, H1, and H2) to the process model in Figure 1. Such redundancy causes confusion for process analysts and has a negative impact on the simplicity and comprehension of the discovered process model.

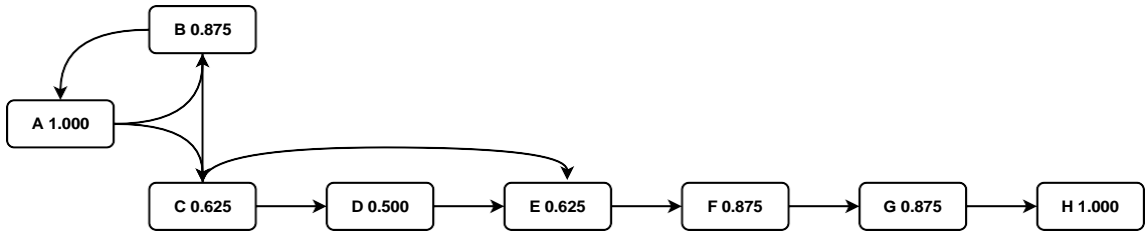

**Figure 1.** Process model discovered from clean event log $L_1$.

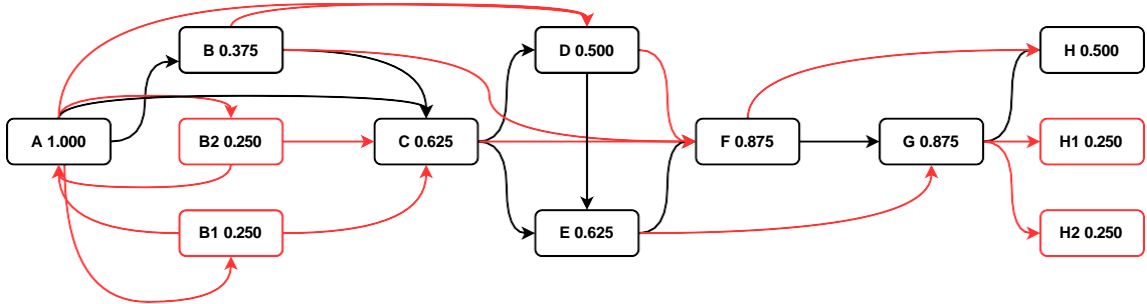

**Figure 2.** Process model discovered from redundant event log $L_2$.

The goal of this paper is to propose a method as a data preprocessing tool that can accurately detect the redundant activity labels to enhance the quality of event logs for better process analysis in process mining.

### 1.2. Contributions of This Paper

The contributions of this paper are as follows:

- For the purpose of improving the quality of event logs, a novel data preprocessing approach is proposed for process mining.
- A multi-view framework is proposed to detect redundant activity labels in event logs. Especially, our method integrates control–flow relations, attribute values and label semantic information in event logs. In terms of the control–flow relation (i.e., the ordering of activities), we adopt the Earth Mover's Distance (EMD) statistical method to compare the directly follows and indirectly follows relations of different activity labels. In terms of the attribute value (i.e., categorical or numerical values of recorded activity labels), activity labels are firstly clustered and followed by EMD to compare the value's distribution. We assess labels' semantic similarity using the pre-trained NLP model as another view. A consensus is guided by a decision-making mechanism to integrate the results produced from multiple views.
- Experiments on publicly available datasets under various settings show that our approach can accurately detect redundant activity labels even when the redundant activity labels are infrequent and contain numerical values as attributes compared with the existing state-of-the-art approach.
- A case study in the healthcare domain using the 5-year EMR dataset collected from two Local Health Districts (LHDs) in Sydney, Australia [18], further demonstrates that our approach can be used as a preprocessing tool in real-life event logs.

The paper is structured as follows. Section 2 discusses the background. Section 3 introduces the basic concepts used throughout the paper. Our proposed method is presented in Section 4. Section 5 presents the experimental results. A real-life case study is explored in Section 6. The paper concludes with Section 7.

## 2. Related Work

### 2.1. Process Discovery Algorithms

Various process discovery algorithms have been proposed in the last decade to discover accurate process models from inputted event logs. The alpha miner [19] was the very first discovery algorithm that aims to automatically discover Petri nets based on clean and noise-free event logs. Later research papers extended the original alpha miner to discover invisible tasks [3] and non-free-choice behaviours [20]. However, the alpha algorithm suffers from poor quality in real-life event logs. The Heuristic Miner [6] was proposed to handle noisy event logs by calculating the relative frequency of each dependency from event logs and removing the dependency from the directly follows graph based on a predefined threshold. The Inductive Miner [21] adopts a divide-and-conquer approach to recursively filter out infrequent relations from discovered process trees and can discover sound process models. The BPMN Miner [22] can also handle noisy event logs by employing approximate dependency discovery techniques, which enables the algorithm to detect and remove infrequent relations even if not all instances are matched in event logs. The newly proposed split miner [7] utilises the breadth-first forward exploration to search for the best incoming and outgoing edges for each node on the directly follows graph while maintaining its connectivity. The algorithm preserves the most frequent incoming and outgoing relations for each activity and filters out the rest. There are also other types of algorithms for discovering process models, such as the genetic algorithm [23] and declarative algorithms using the association rule mining [4].

To summarise, efforts have been made into proposing algorithms that can accurately discover process models from event logs. However, most process discovery algorithms require high-quality event logs as input. The performance of existing process discovery algorithms can be affected if data quality issues are presented in event logs, such as the redundancy of activity labels.

### 2.2. Event Log Quality

Event log quality has been identified as a critical issue that affects process mining results from many domains in both process discovery and enhancement [1,5]. The process mining manifesto [9] has emphasised the importance of event log quality. The first guideline for process mining is to treat event data as first-class citizens. Later on, Suriadi et al. [10] outlined 11 common event log imperfection patterns, such as incorrect timestamps and redundant labels. Fox et al. [24] and Mans et al. [25] suggested quality frameworks for assessing EMR data in the healthcare domain. Additionally, Bose et al. [26] and Aalst [27] raised concerns for event data quality in process mining. Therefore, apart from proposing advanced process discovery, it is also essential to address data quality as early as the event log level.

However, compared with process discovery, less effort has been made to improve event data quality. Event log quality can be improved by detecting erroneous data and potentially repairing it by relying on a reference model or observed correct data [28]. Conforti et al. [28] proposed to identify and repair events with the same timestamp using information from correctly ordered events in the log. Rogge-Solti et al. designed an approach to identify and restore missing events using a reference process model annotated with execution times. Similarly, Sim et al. [29] proposed likelihood-based multiple imputations by event chains to repair missing events in logs. Alharbi et al. [30] proposed an interval-based event selection method to reduce variations in event logs by targeting the behaviour of repeated activities.

In order to detect redundant activity labels, relevant works [11,13,14] suggest ways to address this issue at the event log level. Sadeghianasl et al. [11] proposed a contextual approach that takes control–flow relations, resources, time, and data attributes into consideration. From the control-flow perspective, the method reports the similarity between rows of the footprint matrix, which may not well distinguish the frequency difference between two activity labels and suffers from noisy or infrequent relations. Thus, the method achieves relatively low accuracy with low-occurrence-frequency activity labels. Other methods largely adopt a Probability Density Function to assess the value distributions between activity labels. This approach reports relatively poor results if there are numerical values as activity attributes, and this is a common phenomenon in a real-life event log. For instance, healthcare logs contain various laboratory tests and medications as activity labels. It relies on a weighted clustering method to combine the final results, which requires domain knowledge or ground truth to determine the best weight setting. The other method [13,14] collaboratively and interactively detects problematic activity labels by adopting a gamified crowdsourcing approach, which utilises gamification elements (e.g., badges) to encourage a large group of domain experts to identify and repair redundant activity labels.

The issue of activity labels has also been studied in process matching areas at the model level [31–34]. These approaches match two process models from different data sources with the aim to find similar structures and activity labels. It is difficult to address redundant labels within the same log since separated logs may have incomplete processes. Hence, they are more widely used in a process similarity comparison instead of solving problematic event logs. Other approaches [35,36] look at activity labels themselves while ignoring other information from logs, which may cause erroneous results.

Though previous studies have made efforts to address redundant activity labels [11,13,14], many of them have difficulties identifying activity labels with low-occurrence frequencies, or if invalid labels have been used. However, redundant labels usually occupy a small portion of event logs. Many of these approaches rely on event logs with categorical resources as attributes instead of numerical values [11]. Nevertheless, many activities have such attributes, especially in healthcare logs, such as laboratory tests and observations. Other approaches [13,14] require domain knowledge to improve the data quality. Hence, we propose a multi-view framework to consider contextual information from event logs and aim to accurately detect redundant activity labels without domain knowledge even when the redundant activity labels are infrequent and contain numerical values as attributes.

## 3. Preliminaries

### 3.1. Problem Definition

In this section, we introduce some basic concepts used in this paper.

**Definition 1** (Event Log, Trace, Activity, Event). *An event log $L$ is a collection of traces. A trace $t \in L$ is a sequence of events. $A$ is the set of activities, and an event $e$ is an execution record of an activity $a \in A$. $\#_a(e)$ is a function that obtains attribute values recorded for an event $e$. For example, $\#_{activity}(e)$ obtains the activity name for an event, and $\#_{attribute}(e)$ obtains the attribute value for an event.*

For example, let $E = \{a, b, c, d\}$ be a set of activities. $t = <e_1, e_2>$ is a trace, where $\#_{activity}(e_1) = a$ and $\#_{activity}(e_2) = b$. $L = \{<e_1, e_2>, <e_3, e_4>\}$ is an event log, where each $e_n$ represents a unique execution record of a specific activity. For the sake of understanding, $\#_{activity}(e_n)$ will be used for each event for the rest of the paper.

**Definition 2** (Ordering Relation). *Let $L$ be an event log and $t \in T$ be a trace. For $a, b \in A$, the ordering relations between $a$ and $b$ are defined as follows:*

- *Directly follows relation: $a >_W b$ holds if there is a trace $t = <e_1, e_2, e_3, e_4, ..., e_n>$ and $i \in \{1, 2, 3, ..., n-1\}$ such that $t \in L$ and $\#_{activity}(e_i) = a$ and $\#_{activity}(e_{i+1}) = b$.*
- *Indirectly follows relation: $a >>_W b$ holds if there is a trace $t = <e_1, e_2, e_3, e_4, ..., e_n>$ and $i < j$ and $i, j \in \{1, 2, 3, ..., n-2\}$ such that $t \in L$ and $\#_{activity}(e_i) = a$ and $\#_{activity}(e_j) = b$.*

Since we are only interested in indirectly follows relations with strong connections, we define the following measurement based on [6] to measure how reliable an indirectly follows relation is.

**Definition 3** (Long Distance Measure).

$$a \Rightarrow_W b = \frac{\left(2(|a \gg_W b|)\right)}{|a| + |b| + 1} - \frac{\left(2Abs(|a| - |b|)\right)}{|a| + |b| + 1} \qquad (1)$$

If the first part of the equation is close to 1, then activity $a$ is always followed by activity $b$. The second part of the equation measures the frequency distributions for activities $a$ and $b$. A value close to 0 indicates that the frequency of activities $a$ and $b$ is about equal [6]. Therefore, a value of the long-distance measure close to 1 means that the activities $a$ and $b$ have a strong indirectly follows relation. Based on this, we define a strong indirectly follows relation between two activities.

**Definition 4** (Strong Indirectly Follows Relation). *$a \ggg_W b$ holds between two activities $a, b \in A$ if $a \Rightarrow_W b$ is larger than a given threshold $p$. In this paper, $p$ is set to 0.9, as recommended in [6].*

For example, $L = \{< a, b, c, d >, < b, c, d >\}$ is an event log. The direct-follows relations are $a >_W b$, $b >_W c$, and $c >_W d$. The indirectly follows relations are $a \gg_W c$, $a \gg_W d$, and $b \gg_W d$. The long distance measures for these three indirectly follows relations are 0, 0, and 0.92. Thus, the strong indirectly follows relation is $b \ggg_W d$.

**Definition 5** (Directly Follows Graph). *A directly follows graph is defined as $G = (A, K)$ where $A$ is a finite set of activities in the event log (same as Definition 1), and $K \subseteq A \times A$ is a set of directed arcs, which represent directly follows relations (i.e. $a >_W b$ exists if $(a, b) \in K$). An example is shown in Figure 3.*

**Definition 6** (Pre-Sets and Post-Sets). *Let $G = (A, K)$ be a directly follows graph. For $a \in A$, we have*
*$a \bullet = \{b | b \in A \land (a, b) \in K\}$ and*
*$\bullet a = \{b | b \in A \land (b, a) \in K\}$,*
*where $a \bullet$ and $\bullet a$ are called a post-set and a pre-set of $a$, respectively. $a \bullet$ represents all the directly outgoing activities from $a$, e.g. $A \bullet = \{H, B\}$, and $\bullet a$ represents all directly incoming activities to $a$, e.g. $\bullet C = \{H, B\}$, as shown in Figure 3.*

**Definition 7** (Count Frequency). *$|(a, b)|, (a, b) \in K$ counts how many times the relation $a >_W b$ occurs in $G$ (e.g., $|(A, H)| = 50$ in Figure 3).*

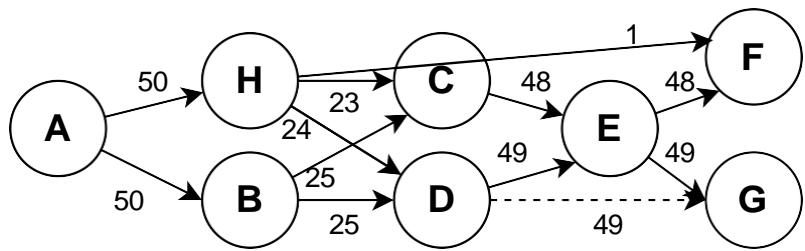

**Figure 3.** An example direct-follows graph.

## 4. A Multi-View Detecting Framework

This section describes our proposed framework, shown in Figure 4, to detect redundant activity labels. The underlying principle is that redundant activity labels share the same patterns on both control–flow relations and attribute values. We also include semantic similarity as an additional view. Thus, our approach assesses similarities from the above views using an Earth Mover's Distance (EMD) statistical method and a pre-trained NLP model. To this end, we first introduce EMD to compare control–flow relations probability distributions. We then demonstrate how to extend EMD to calculate attribute value similarity. We apply a powerful NLP model in semantic similarity. Finally, we briefly describe how to use the decision-making mechanism to combine results from different views to obtain the final output.

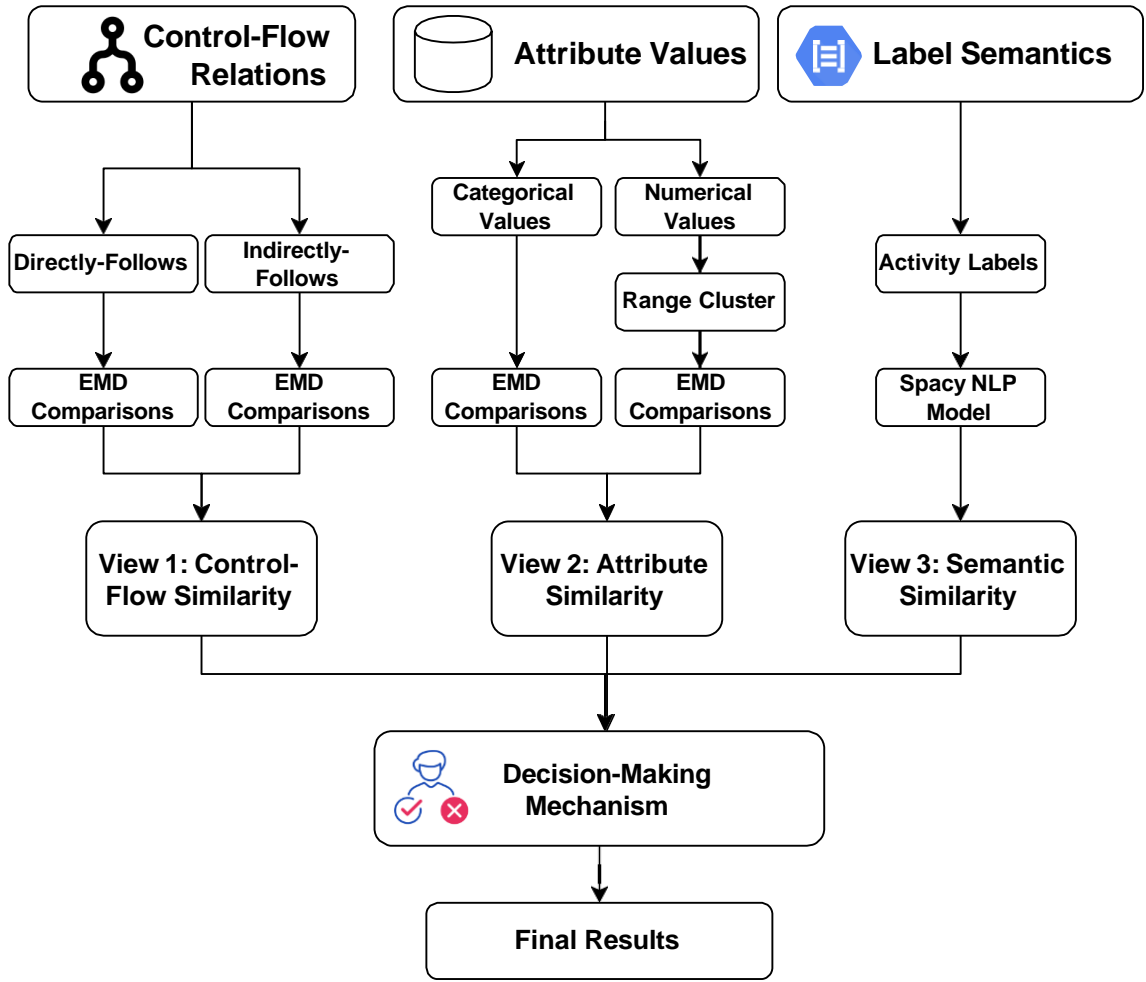

**Figure 4.** Overview of the proposed approach.

### 4.1. Earth Mover's Distance

The Earth Mover's Distance (EMD) [37] is a method for comparing two multiple dimensional probability distributions over a region. It was first proposed as a matrix to retrieve images in the computer vision domain. However, it has been applied to many other fields [38,39]. The EMD calculates the lowest costs of transferring one distribution into another, given that two distributions indicate different ways of accumulating a certain amount of dirt in a region. A distance function defines the cost needed to move dirt between certain piles. It is frequency-aware, as it considers the magnitude of discovered differences, and the difference is determined by the ground distance function that can express different perceptions of similarity [40]. Below we formally introduce the EMD:

Let P be a probability distribution with $p_1, ..., p_m \in P$ as different clusters and $w_{p1}, ...w_{pm} \in \mathbb{R}^+$ as the associated weight for these clusters. Another probability distribution Q has the same notations $(q_1, w_{q1}), ...(q_n, w_{qn})$. A ground distance $D = d(p_i, q_j)$ between cluster $p_i$ and $p_j$ is defined. We seek to find a flow $F = (f_{i,j}) \in \mathbb{R}^{m \times n}$ that minimises the overall costs to transfer P to Q. The following constraints should be followed:

- Non-negativity flow: $f_{i,j} \geq 0, \forall 1 \leq i \leq m, 1 \leq j \leq n$.
- Sent and receive flow should not exceed weights in P and Q:

- $\sum_{j=1}^{n} f_{i,j} \leq w_{pi}, \forall 1 \leq j \leq n;$
- $\sum_{i=1}^{m} f_{i,j} \leq w_{qj}, \forall 1 \leq i \leq m.$

- All weights possible have to be sent: $\sum_{i=1}^{m} \sum_{j=1}^{n} f_{i,j} = min(\sum_{i=1}^{m} w_{pi}, \sum_{j=1}^{n} w_{qj})$

The optimal flow F is defined as

$$EMD(P, Q) = min \frac{\sum_{i=1}^{m} \sum_{j=1}^{n} f_{i,j} d(p_1, q_j)}{\sum_{i=1}^{m} \sum_{j=1}^{n} f_{i,j}} \qquad (2)$$

### 4.2. View 1: Measuring Control–Flow Similarity

This section introduces the approach used to calculate similarity for the control–flow relations of activity labels. Redundant activity labels should share similar control–flow relations or ordering patterns. This similarity indicates not only identical control–flow relations but also closed distribution patterns. As shown in Figure 4, the overall idea behind a control–flow view is that, for each pair of activity labels $(a_i, a_j) \in A$, we adopt EMD to compare the directly follows and the strong indirectly follows relations along with their distributions. Each directly follows and strong indirectly follows comparison can be further divided into directly and indirectly outgoing (i.e., consequence) and incoming (i.e., precedent) relations. Thus, we obtain four different values, with the final similarity being the weighted average of these values.

The control–flow view is separated by directly follows and strong indirectly follows comparisons. We would like to place most of our effort on explaining the directly follows comparison, since the strong indirectly follows comparison follows the same algorithm, only the relations are strong indirectly follows. The reason we also consider strong indirectly follows relations is to handle non-free-choice problems (i.e., whether we choose a task is dependent on what has been executed in the prior process [41]). For instance, both activities $C$ and $D$ in Figure 3 have identical directly follows relations, but $D >>>_W G$ (i.e., the dashed line) also exists. Thus, $C$ and $D$ should not be regarded as redundant.

Algorithm 1 presents our approach for calculating the directly follows similarity. The starting point is to construct a directly follows graph obtained from the event log (Line 1). For each activity label, we then calculate its outgoing and incoming activity sets (Line 2-4). By using Equation (2), the weights are calculated for each element in the activity set (Line 5-6), (e.g., $A \bullet_W = \{\frac{1}{2}, \frac{1}{2}\}$). Afterwards, for each pair of activity labels, we adopt EMD to calculate the similarity between incoming and outgoing activity sets using the ground distance function $D_{cf}$ from Equation (3) (Line 7-9). The activities in the sets (e.g., $a_\bullet$) are clusters. The weights in the sets (e.g., $a_{\bullet W}$) are the associated weights for each cluster. For instance, suppose we would like to calculate the similarity between outgoing activity sets for $H$ and $B$ in Figure 3, the input signatures for EMD would be $P = \{(C, 0.46), (D, 0.48), (F, 0.02)\}$ and $Q = \{(C, 0.5), (D, 0.5)\}$ Lastly, the directly incoming and outgoing similarities are averaged to obtain the final directly follows similarity for each pair of activity labels and added to the set $S_d$ (Line 10-11).

The equation for calculating the weight of a single activity in incoming/outgoing activity sets is defined as

$$W = \frac{|(b, a)|}{\sum\{|(\epsilon, a)|\epsilon \in \bullet a\}} \qquad or \qquad \frac{|(a, b)|}{\sum\{|(a, \epsilon)|\epsilon \in a \bullet\}} \qquad (3)$$

The ground distance function $D_{cf}$ for EMD between any two clusters $p_i, q_j$ from activity sets is defined as

$$D_f = \begin{array}{ll} 0 & \text{if } p_i = q_j \\ 1 & \text{otherwise} \end{array} \qquad (4)$$

*Principle.* The same activity label has no cost, and different ones have a unit cost. This cost function can be easily extended based on other matrices, e.g., global location for activities. Here, we just show the most basic version for better undesirability.

---

**Algorithm 1:** Directly Follows Similarity

---

**Input:** Event log $L$
**Output:** $S_d$: Set of Directly Similarities for All Pairs of Activities

1  $G \leftarrow$ MakeDirectlyFollowsGraph($L$);
2  **foreach** $a \in A$ **do**
3      $a \bullet \leftarrow$ OutGoing($a$);
4      $\bullet a \leftarrow$ InComing($a$);
5      $a \bullet_W \leftarrow$ CalculateWeight($a \bullet$);
6      $\bullet a_W \leftarrow$ CalculateWeight($\bullet a$);
7  **foreach** $a, b \in A$ **do**
8      Outgoing Similarity = EMD($a \bullet$, $a \bullet_W$, $b \bullet$, $b \bullet_W$, $D_{cf}$);
9      Incoming Similarity = EMD($\bullet a$, $\bullet a_W$, $\bullet b$, $\bullet b_W$, $D_{cf}$);
10     Directly Follows Similarity = Average(Outgoing Similarity, Incoming
        Similarity);
11     $S_d \leftarrow$ Directly Follows Similarity;
12 **return** $S_d$

---

The same algorithm applies to the calculation of indirectly incoming and outgoing similarities. We construct a strong indirectly follows graph from the event log. We have a set that contains strong indirectly follows similarities as well. Thus, for each pair of activity labels, the overall control–flow similarity is the weighted average of directly and strong indirectly follows similarities, where there is a value between 0 and 1. The greater the value is, the more significant the effort needed to transfer one distribution to another is, which means that the two activity labels have less similarity with regard to the control–flow perspective. The combination of four different scores can be easily extended with other statistical or clustering algorithms, e.g., k-means clustering. We seek to show that our approach can achieve desirable results with the most fundamental method and requires no domain knowledge, e.g., the number of clusters, as the input.

### 4.3. View 2: Measuring Attribute Value Similarity

This section introduces the approach used to calculate similarity for the attribute values of activity labels. In an event log, attribute values can be the resources needed for executions (e.g., the person who performed the event) or the associated recorded values when executing the event (e.g., the result value for the event). Redundant activity labels share the same attribute values. Therefore, the proposed framework should incorporate attribute values (i.e., both categorical and numerical values) when detecting redundant activity labels.

The overall approach, shown in Figure 4, can be divided into two parts: activity labels with categorical values and numerical values. Activity labels with categorical values are relatively easy to calculate. We calculate the frequency distribution of the attribute values for each activity label and apply pairwise EMD between different activity labels. The ground distance function borrows the idea from Equation (4): the same attribute value has no cost, and different ones have a unit cost. For activity labels with numerical values, we firstly cluster each activity label into different clusters based on value percentiles. We then apply EMD to assess data distributions of activity labels within each cluster. Clustering first ensures that only activity labels with the same data range are further evaluated. Activities with different data ranges are unlikely to share similar data patterns, which do not need to be further assessed for data distributions.

We describe how our approach calculates attribute value similarity for activity labels with numerical values in Algorithm 2. For each activity, we first assess whether this activity has a numerical value attribute (Line 2). If not, it has minimal attribute value similarity with other activity labels (i.e., *AttributeValueSimilarity* = 1). If yes, Line 3 finds all events of the activity ($\#_{activity}(e) = a$) and obtains numerical values for that attribute ($\#_{attribute}(e)$) into a dataset (i.e., *DataSet$_a$*). Line 4 calculates the 25th and 75th percentiles for each dataset. We use the 25th and 75th percentiles as a 2-D vector and apply Agglomerative Hierarchical Clustering [42] with a threshold $\theta_a$ for all datasets (Line 5). For example, if an activity label has the values 0.3 and 0.5 for its 25th and 75th data value percentiles, respectively, then the 2-D vector is (0.3, 0.5). We apply the Euclidean Distance [43] as the distance measurement between two vectors. Activity labels that are not in the same cluster also have *AttributeValueSimilarity* = 1. There are many unique values in the attribute; it is hard to directly apply EMD because of the many different clusters in the distribution. As a result, we transfer each dataset to a histogram following Sturges' formula [44], where uniform maximum and minimum values are used to ensure two histograms have the same bin number and size when comparing activity label pairs within the same cluster (Line 11). We pick each interval's left boundary as cluster values (e.g., $p_i$, $q_j$), and Line 12-13 calculate the percentage of each bin as cluster weights (e.g., $w_{pi}$, $w_{qj}$). An example cluster is (10, 20%), (15, 30%), (20, 20%), and (25, 30%). In this way, we transfer each histogram as a cluster, and EMD is further used to compare two clusters using the distance function $D_d$ in Equation (5) (Line 14). The *AttributeValueSimilarity* is normalised [? ] to become a value between 0 and 1 and added to $S_n$. Similar to the control–flow perspective, in the attribute value perspective, the greater the value is, the less similarity they have.

---

**Algorithm 2:** Attribute Value Similarity

---

**Input:** Event log $L$, threshold $\theta_a$
**Output:** $S_n$: Set of Attribute Value Similarities for all Pairs of Activities

1 **foreach** $a \in A$ **do**
2      **if** *HasNumericalValueAttribute(a)* **then**
3          $DataSet_a$ ← ExtractAttribute($a$);
4          $Q_{1a}$, $Q_{3a}$ ← CalculatePercentiles($a$);
5 Clusters ← AgglomerativeHierarchicalClustering($Q_{1a}$, $Q_{3a}$, $\theta_a$);
6 **foreach** $C \in Clusters$ **do**
7      **if** *size(C)* $\leq 1$ **then**
8          continue;
9      **else**
10          **foreach** $a, b \in C$ **do**
11              $H_a$, $H_b$ ← MakeHistograms($DataSet_a$,   $DataSet_b$);
12              $H_{aW}$ ← CalculateWeight($H_a$);
13              $H_{bW}$ ← CalculateWeight($H_b$);
14              Attribute Value Similarity = EMD($H_a$, $H_{aW}$, $H_b$, $H_{bW}$, $D_d$);
15              $S_n$ ← Attribute Value Similarity;
16 **return** $S_n$

---

The ground distance function $D_d$ for EMD between any two attribute clusters $p_i$, $q_j$ from histograms is defined as

$$D_d = |p_j - q_j| \tag{5}$$

*Principle.* Since both $p_i$, $q_j$ are numerical values, it takes less effort to transfer $p_i$ to $q_j$ if they are close to each other. We adopt the difference between $p_i$ and $q_j$ as the ground distance function.

### 4.4. View 3: Measuring Semantic Similarity

We are interested in measuring the semantic similarity between activity labels, since redundant activity labels may share close semantics. Although simply looking at the semantics of activity labels may lead to false-positive results (i.e., labels that are incorrectly detected as redundant), it is still an important factor to consider when detecting redundant labels. There are many methods proposed for assessing semantic similarity between words, such as the string edit distance [45]. However, the string edit distance cannot easily handle synonymous activity labels with a different wording structure. Therefore, we integrate NLP into our multi-view framework. We apply a pre-trained NLP model Spacy, which is a well-known industrial-strength NLP tool that provides fast and accurate syntactic analyses, to assess the semantic similarity between every pair of activity labels [46]. The result between a pair of activity labels is a numerical number ranging from 0 to 1, where 1 means that they are identical and vice versa. However, in order to comply with the same rule setting in previous sections, we subtract the obtained results from 1. In this way, 0 means that they are identical in semantic similarity

### 4.5. Decision-Making Mechanism: Majority Voting

For now, each pair of activity labels has three features of similarities: control–flow relations, attribute values, and label semantics. This section describes a decision-making mechanism to aggregate similarities from the above three features and generates final results, as shown in Figure 4. The decision-making mechanism is the majority voting, which is a widely adopted concept in ensemble learning [47]. For majority voting, we first need to determine the threshold for each feature (i.e., $\theta_c$, $\theta_d$, and $\theta_s$) to decide whether each pair of activity labels is similar in the corresponding feature. Equation (6) describes the voting mechanism, where $V_i$ represents the result from a particular view, and $m$ represents the total number of views. The activity labels are detected as redundant if more than half of the views are regarded as similar and vice versa. Using the majority voting as the decision-making mechanism has the following advantages: (1) it is fast and easy to implement, and requires no domain knowledge as input, (2) it can be easily extended if more views are proposed for redundant activity label detection, and (3) it can be easily integrated with domain knowledge (e.g., the voting weight of each view).

$$Result = \begin{cases} \text{``Redundant''} & \text{if } |i : V_i = \text{``Similar''}| \geq m/2 \\ \text{``Non-redundant''} & \text{otherwise} \end{cases} \quad (6)$$

## 5. Evaluation

We conducted a large number of experiments to prove that our proposed framework can accurately detect redundant activity labels in event logs. Overall, two groups of experiments were performed. The first group of experiments compared the performance of our approach with the existing state-of-the-art method to detect redundant activity labels. The second group of experiments further analysed the effectiveness of our method.

To evaluate the performance of our method, we apply the same evaluation matrix as found in [11], which is the standard f-score metric. Detection results fall into one of the following four categories: (1) true-positive (TP), where positive outcomes are correctly detected (the detected redundant activity label is actual redundant), (2) false-positive (FP), where negative outcomes are detected as positive (the detected redundant activity label is actual not redundant), (3) true-negative (TN), where negative outcomes are correctly detected (the detected non-redundant activity label is actual not redundant), and (4) false-negative (FN), where negative outcomes are falsely detected (the actual redundant activity label is detected as non-redundant). Based on these four indicators, we calculate the following evaluation scores. Precision defines how many positive classes are detected correctly out of all positive detection.

$$Precision = \frac{TP}{TP + FP} \quad (7)$$

Recall indicates how many from all positive classes are detected correctly.

$$Recall = \frac{TP}{TP + FN} \quad (8)$$

The f-score is the harmonic mean of precision and recall.

$$F\_score = \frac{2 \times Precision \times Recall}{Precision + Recall} \tag{9}$$

Several publicly available datasets were utilised in the experiments. In total, our evaluation is based on four publicly available event logs:

- Hospital Billing log[1]: An event log records processes related to billing medical services provided by a Dutch hospital.
- Sepsis log[2]: An event log records treatment processes of sepsis patients from a Dutch hospital.
- Helpdesk[3]: An event log contains the ticketing management process of the help desk in a software company in Italy.
- BPI Challenge 2012[4]: An event log of a loan application process in a Dutch financial institute.

The details of all used event logs are presented in Table 1. Following the data preparation methods in [11], we randomly selected a certain amount of activity labels and randomly varied the percentage (i.e., 1% to 30%) of its events to simulate activity labels with both low-occurrence and high-occurrence frequency. In total, seven different settings were used on four event logs. For each setting, 5 rounds were performed, and the average results are reported. In total, 70 event logs were generated and evaluated in the experiment. It is worth mentioning that the Sepsis event log also contains different variants of discharging a patient, which are "*Release C*", "*Release D*", and "*Release E*". They are regarded as redundant [11,17]. Thus, the ground truth not only contains activity labels that we manually changed, but also consists of any pair of these three activity labels.

In real-life situations, activity labels that are similar in any two aspects will be regarded as redundant according to the majority voting mechanism. Since the data quality on the activity label is a unique problem in process mining, even though the redundant activity labels are similar in semantics, they still need to be regarded as similar in terms of the control–flow relation or the attribute value. Hence, we are more interested in evaluating whether our approach can detect redundant activity labels that are different in semantics, because the effectiveness of the first two features is critical in the proposed framework. Therefore, we artificially renamed the activity labels arbitrarily following [11] and intentionally set the voting weight of semantic similarity to 0 when performing the experiments. Hence, activity labels that are similar in both two aspects were regarded as redundant in the experiments. We will further demonstrate the use of semantic similarity in the real-life case study proposed in Section 6. The proposed framework was implemented as a Python program for evaluations. We adopted $\theta_c$ =0.25 and $\theta_d$ =0.1 for each aspect.

**Table 1.** Characteristics of event logs used for evaluations.

| Event Log | Number of Traces | Number of Trace Variants | Number of Events | Number of Attributes | Number of Activity Labels |
|-----------|-----------------|-------------------------|------------------|---------------------|---------------------------|
| Hospital Billing | 100,000 | 1,020 | 451,359 | 1,105 | 18 |
| Sepsis | 1,050 | 846 | 15,214 | 26 | 16 |
| Helpdesk | 4,580 | 226 | 21,348 | 22 | 14 |
| BPI Challenge 2012 | 13,087 | 4,366 | 262,200 | 69 | 24 |

### 5.1. Comparing with The Existing Method

Currently, three state-of-the-art approaches are proposed for detecting redundant activity labels in event logs [11,13,14]. However, the approaches in [13,14] are an interactive detection approach, which requires domain experts. Hence, for the baseline approach, we selected the SynonymousLabelRepair [11], which seems to be more advanced in handling redundant activity labels and requires less domain knowledge than the other methods. We used its default settings in evaluations, which are a 0.7 threshold and a uniform weight, as the optimal weight settings required extra domain knowledge.

We followed the evaluation pipeline proposed in [11] for the first group of experiments, which aimed at comparing the performance of our proposed framework with the existing method to detect redundant activity labels in event logs, as shown in Figure 5. The "Hospital Billing" and "Sepsis" event logs were evaluated by SynonymousLabelRepair in [11]. Hence, in this section, we compare our approach with the SynonymousLabelRepair using the above two event logs.

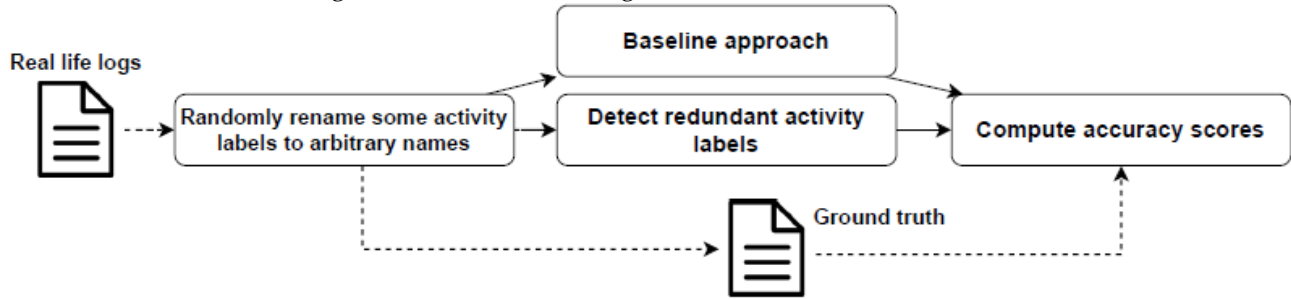

**Figure 5.** Overview of the evaluation pipeline.

Table 2 along with Figure 6 and 7 show the results for the evaluation. It is clear that our approach outperforms the baseline approach. The f-scores of our approach are all above 0.8 for the two event logs, which indicates that almost all redundant activity labels can be successfully detected regardless of their occurrence frequencies in event logs. On the contrary, the average f-scores for the baseline were merely 0.33 and 0.43 in the experiments. Moreover, the baseline performs poorly when the redundant activity labels are less frequent. For instance, the baseline only achieves a 0.08 f-score when there are 1% redundant activity labels in the "Hospital Billing" event log. We also noticed that the baseline begins to catch up and even surpasses our approach in recall when redundant activity labels become more frequent in the "Hospital Billing" event log, which shows that both approaches can successfully detect most of the redundant activity labels. However, the low precision values indicate that the baseline constantly produces false-positive results (i.e., non-redundant activity labels are falsely detected as redundant), which further proves the necessity and effectiveness of our multi-view approach.

The baseline approach only compares directly follows relations while ignoring their frequency distributions and indirectly follows relations. However, low-frequency activity labels rarely contain all directly follows relations while only maintaining the main one. In this case, frequency distributions of the control–flow relations are becoming important in the approach. We also notice that the control–flow view is limited when handling activity labels that share XOR relations, such that they have identical incoming and outgoing relations. This further illustrates the necessity to also consider other views when detecting the redundant activity labels. Furthermore, the cost function between different activity labels in EMD can be better defined instead of simply adopting the unit cost if domain knowledge is available. Thus, more satisfying results can be potentially achieved by our approach. Additionally, the baseline relies on a Probability Density Function to assess value distributions for activity attributes. However, distributions of attribute values are less structured when redundant activity labels are infrequent and mixed with categorical and numerical values.

**Table 2.** Comparison of our method with SynonymousLabelRepair [11].

| Event Log | Number of Redundant Activity Labels | Precision | | Recall | | F-score | |
| --- | --- | --- | --- | --- | --- | --- | --- |
| | | **Ours** | **Baseline** | **Ours** | **Baseline** | **Ours** | **Baseline** |
| Hospital Billing | 5,254 (1%) | 0.97 | 0.05 | 0.89 | 0.20 | 0.93 | 0.08 |
| | 21,693 (5%) | 0.94 | 0.17 | 0.90 | 0.73 | 0.92 | 0.28 |
| | 44,890 (10%) | 0.88 | 0.5 | 0.80 | 0.18 | 0.84 | 0.26 |
| | 66,368 (15%) | 0.85 | 0.19 | 0.89 | 0.90 | 0.87 | 0.31 |
| | 90,273 (20%) | 0.87 | 0.24 | 0.86 | 0.92 | 0.86 | 0.38 |
| | 112,840 (25%) | 0.80 | 0.30 | 0.94 | 0.85 | 0.86 | 0.44 |
| | 135,480 (30%) | 0.80 | 0.42 | 0.96 | 0.95 | 0.87 | 0.58 |
| Sepsis | 180 (1%) | 0.76 | 0.39 | 0.90 | 0.23 | 0.82 | 0.29 |
| | 745 (5%) | 0.75 | 0.47 | 0.89 | 0.42 | 0.81 | 0.44 |
| | 1,569 (10%) | 0.93 | 0.52 | 0.77 | 0.45 | 0.84 | 0.47 |
| | 2,327 (15%) | 0.93 | 0.33 | 0.71 | 0.25 | 0.81 | 0.29 |
| | 3,086 (20%) | 0.90 | 0.48 | 0.76 | 0.46 | 0.83 | 0.47 |
| | 3,844 (25%) | 0.80 | 0.55 | 0.85 | 0.49 | 0.82 | 0.51 |
| | 4,605 (30%) | 0.86 | 0.52 | 0.77 | 0.58 | 0.82 | 0.55 |

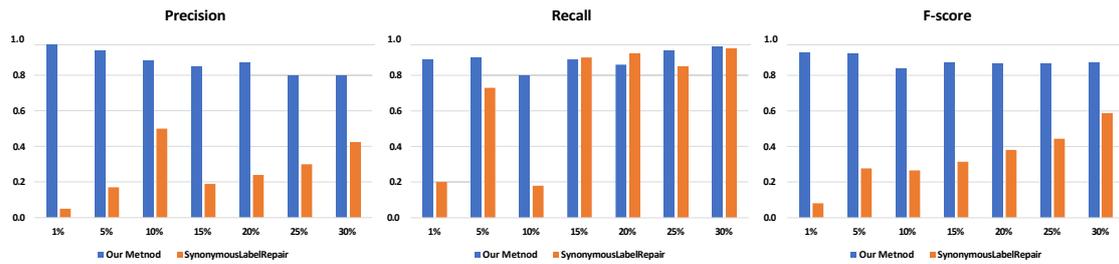

**Figure 6.** Comparison of our method with the baseline using the Hospital Billing event log.

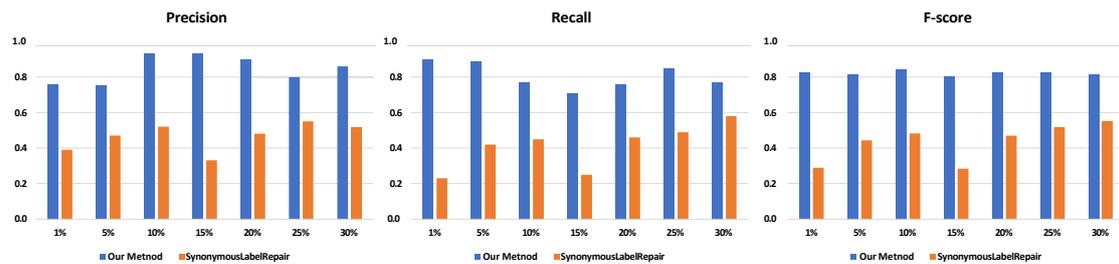

**Figure 7.** Comparison of our method with the baseline using the Sepsis event log.

*5.2. Further Analysis of Our Proposed Method*

To further analyze the performance of the proposed approach, more evaluations are performed with more event logs and redundant activity labels. Apart from running experiments on our multi-view approach, we also implemented two baselines to prove the effectiveness of our approach:

- Control–Flow Only: the baseline only relies on the control-flow similarity to detect redundant activity labels.
- Attribute Value Only: the baseline only relies on the attribute similarity to detect redundant activity labels.

In total, all four event logs are used to evaluate our approach. For each event log, we randomly renaming 1%, 5%, 10%, 15%, 20%, 25% and 30% of the activity labels. Table 3 and Figure 8 present the f-score comparisons between the baselines and our approach. All f-scores are averaged by 5 repeats. In total, 140 event logs are generated and evaluated in the experiment. The f-scores of our approach are much higher than the two baselines, which only detect from a single perspective. The results suggest that it is important to consider information from multiple views when detecting redundant activity labels. We also notice that, except for the BPI Challenge 2012 event log, the control–flow view usually plays a more important role than the attribute value view.

**Table 3.** Further analysis of our method.

| Event Log | Number of Redundant Activity Labels | Control–Flow Only | Attribute Value Only | Our Method |
|---|---|---|---|---|
| Hospital Billing | 5,254 (1%) | 0.72 | 0.71 | 0.93 |
| | 21,693 (5%) | 0.74 | 0.69 | 0.92 |
| | 44,890 (10%) | 0.70 | 0.66 | 0.84 |
| | 66,368 (15%) | 0.78 | 0.67 | 0.87 |
| | 90,273 (20%) | 0.71 | 0.66 | 0.86 |
| | 112,840 (25%) | 0.69 | 0.65 | 0.86 |
| | 135,480 (30%) | 0.67 | 0.63 | 0.87 |
| Sepsis | 180 (1%) | 0.66 | 0.60 | 0.82 |
| | 745 (5%) | 0.63 | 0.55 | 0.81 |
| | 1,569 (10%) | 0.70 | 0.67 | 0.84 |
| | 2,327 (15%) | 0.65 | 0.58 | 0.81 |
| | 3,086 (20%) | 0.71 | 0.58 | 0.83 |
| | 3,844 (25%) | 0.69 | 0.66 | 0.82 |
| | 4,605 (30%) | 0.72 | 0.61 | 0.82 |
| Helpdesk | 213 (1%) | 0.73 | 0.66 | 0.92 |
| | 1,067 (5%) | 0.70 | 0.62 | 0.89 |
| | 2,135 (10%) | 0.77 | 0.58 | 0.90 |
| | 3,202 (15%) | 0.71 | 0.60 | 0.87 |
| | 4,270 (20%) | 0.71 | 0.62 | 0.88 |
| | 5,337 (25%) | 0.65 | 0.70 | 0.85 |
| | 6,404 (30%) | 0.68 | 0.65 | 0.84 |
| BPI Challenge 2012 | 2,622 (1%) | 0.58 | 0.65 | 0.86 |
| | 13,110 (5%) | 0.60 | 0.71 | 0.88 |
| | 2,6220 (10%) | 0.65 | 0.63 | 0.83 |
| | 39,330 (15%) | 0.60 | 0.61 | 0.84 |
| | 52,440 (20%) | 0.70 | 0.68 | 0.88 |
| | 65,550 (25%) | 0.68 | 0.66 | 0.85 |
| | 78,660 (30%) | 0.63 | 0.63 | 0.82 |

Finally, we also conducted experiments to show how redundant activity labels can impact the discovered process models and how our approach can be used as a preprocessing tool to repair event logs. We selected two event logs with the highest (Hospital Billing) and lowest (Sepsis Log) detection f-score for the experiments. A simple and fundamental repair mechanism was adopted. Redundant activity labels were replaced with the most similar non-redundant activity labels in event logs. A mode of advanced repair technology is left as future work. We aimed to compare the f-scores for discovered process models on original event logs (no redundant activity labels exist), problematic event logs (with redundant activity labels), and repaired event logs using the proposed detection framework with the basic repair approach. We used the Inductive Miner infrequent [21] to mine process models. We conducted conformance checking between all discovered process models against the original event logs. The more conformance there was between the process model discovered from the repaired event log and the original event log, the higher the f-score achieved. We used alignment-based fitness and conformance-based precision tools in the PM4PY [48]. The f-scores for the original event logs are referenced from [15] directly. The results are presented in Table 4. We can see that the existence of redundant activity labels has a tremendous impact on the discovered process models. For instance, the f-scores drop significantly compared with the original f-scores. On the contrary, the f-scores on process models discovered from repaired event logs drop slower and remain closer to the original event log. The results indicate that our approach can be used as a preprocessing tool and help with improving the quality of process models when there are redundant activity labels in the event logs.

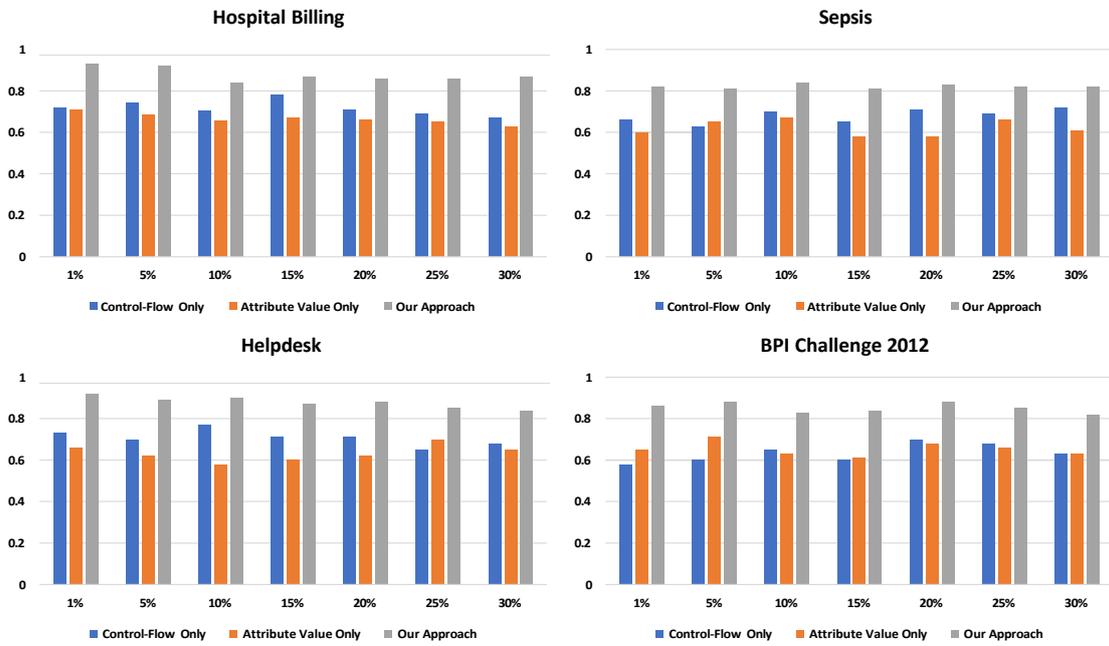

**Figure 8.** Comparison of our method with the baseline using the Hospital Billing event log.

**Table 4.** F-score comparisons between process models discovered from the original, problematic, and repaired event logs.

| Event Log | Number of Redundant Activity Labels | F-Score (the Original Log | Average F-Score (the Logs with Redundant Activity Labels) | Average F-score (the Repaired Logs) |
|---|---|---|---|---|
| Hospital Billing | 5,254 (1%) | 0.75 | 0.69 | 0.73 |
| | 21,693 (5%) | 0.75 | 0.64 | 0.72 |
| | 44,898 (10%) | 0.75 | 0.62 | 0.70 |
| | 66,368 (15%) | 0.75 | 0.59 | 0.69 |
| | 90,273 (20%) | 0.75 | 0.55 | 0.67 |
| | 112,840 (25%) | 0.75 | 0.53 | 0.67 |
| | 135,480 (30%) | 0.75 | 0.51 | 0.66 |
| Sepsis | 180 (1%) | 0.77 | 0.74 | 0.75 |
| | 745 (5%) | 0.77 | 0.72 | 0.74 |
| | 1,569 (10%) | 0.77 | 0.66 | 0.72 |
| | 2,327 (15%) | 0.77 | 0.64 | 0.70 |
| | 3,086 (20%) | 0.77 | 0.61 | 0.70 |
| | 3,844 (25%) | 0.77 | 0.60 | 0.68 |
| | 4,605 (30%) | 0.77 | 0.57 | 0.67 |

## 6. Real-Life Case study

We conducted a case study using the Speed-Extract EMR dataset to demonstrate that our framework can be used in real-life healthcare event logs. The Speed-Extract dataset comprises retrospective data from a historical dataset extracted between 2013 and 2018 from a single Cerner Millennium EMR domain in Sydney, Australia [18]. The Speed-Extract dataset comprises anonymised patients that presented with suspected Acute Coronary Syndrome (ACS) to facilities in Northern Sydney LHD and Central Coast LHD [18]. We aimed to study the treatment process of ST-elevation myocardial infarction (STEMI) patients, which is a type of heart attack that mainly affects the patient's heart's lower chambers [18]. In this paper, we demonstrate how we extract the event log from the Speed-Extract dataset and apply the proposed framework to improve the quality of the event log. We verify and substitute the detected redundant activity labels using the domain knowledge to obtain a more representative event log. In the end, we apply the existing tool to mine the process models from the two event logs. Comparisons between these two models demonstrate that our proposed framework can be used as a preprocessing method for the event log to obtain a more structured and easier to understand process model. The pipeline for the case study is shown in Figure 9.

The data we used in the Speed-Extract dataset includes the following tables: the *Patient-prep* table contains patients' demographics, such as age and gender; the *Medications-mapped* table records prescription orders for each patient; the *Diagnosis-prep* table presents the diagnosis for each patient using the International Classification of Diseases (ICD)-10 codes.

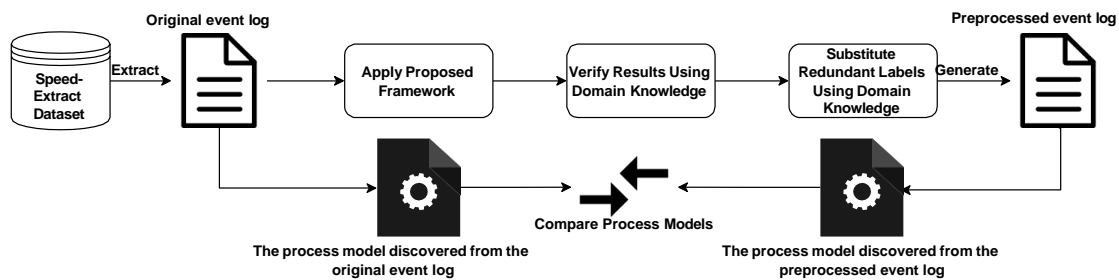

**Figure 9.** Overview of the case study pipeline.

### 6.1. Event Log Construction

STEMI patients were defined using the ICD-10 codes, resulting in 5750 patients. Patients older than 85 or younger than 40 were excluded [49]. We treat patients as traces and their medications as activities to construct the event log. After extracting the initial event log, we filtered out traces where less than one activity was presented, since they exposed less useful information regarding control–flow similarity. As a result, 2141 traces with 615 activities are presented in the event log. The event log has 1363 trace variants due to the high number of activities. The primary reason for such an enormous number of activities is the existence of redundancy in the event log. When giving prescriptions to patients, the free-text nature causes redundancy. For example, when referring to the same medicine, some doctors prefer to write the exact medication name, while others prefer the brand name (e.g., *Telmisartan* is the medication name sold under the brand name *Micardis*). Furthermore, there are many different substitutable medications to choose from when addressing the same symptom, which is one of the other reasons for redundant activity labels in the event log. For instance, both *Candesartan* and *Eprosartan* can be used to treat high blood pressure. Moreover, some doctors prefer giving dose and frequency along with the medication name in one input field, such as *Aspirin one tablet per day*, which also introduces unnecessary redundancy to the event log.

### 6.2. Result and Discussion

Since the goal is to study the treatment process, medications with the same effect shared the same behaviour in the treatment process. Hence, we applied the proposed framework to detect redundancy in the event log and generate a more representative event log for further analysis. In this case study, the thresholds were set to 0.2 for the control–flow relation, 0.1 for the attribute values, and 0.1 for the label semantic information. As a result, nine groups of redundant activity labels were detected according to the transitive property of equality. The results were further evaluated by the domain knowledge to show that these are different therapies in the STEMI treatment [50]. Therefore, we adopted therapy names (e.g., *Beta Blockers*, *Anticoagulants*) to replace redundant activity labels in the original event log to generate the preprocessed event log. Compared with the original event log, the number of activity labels in the process was reduced from about 600 to 13. The number of distinct trace variants was reduced from about 1300 to 660, which is around half of the original event log.

We adopted Disco for process discovery on the event logs [16]. As a result, a "spaghetti-like" process model was discovered from the original event log due to the redundancy. Figure 10 shows a snapshot of the partial process model since the original process model is too complicated to display here. Such a process model cannot provide any useful insights toward understanding the treatment process for STEMI patients. On the contrary, Figure 11 shows the process model discovered from the preprocessed event log. The discovered process model overall seems to be simple and insightful, representing only the important treatment behaviours of STEMI patients. We can observe that most patients were treated with *Aspirin*, *Beta blockers*, and *Statins* during their hospital stays, which is in line with the current treatment guideline [50]. Moreover, several control–flow relations are also presented in the process model, such as *Clopidogrel* $>_W$ *Digoxin* and *Insulin* $>_W$ *Statins*. Such a process model together with additional patients' records can be further used to study other related topics, such as how the treatment process influences a patient's survival time. This case study further shows that our approach can efficiently detect redundant activity labels in the event log. Moreover, the event log with redundant activity labels has a significant impact on the quality of the discovered process model.

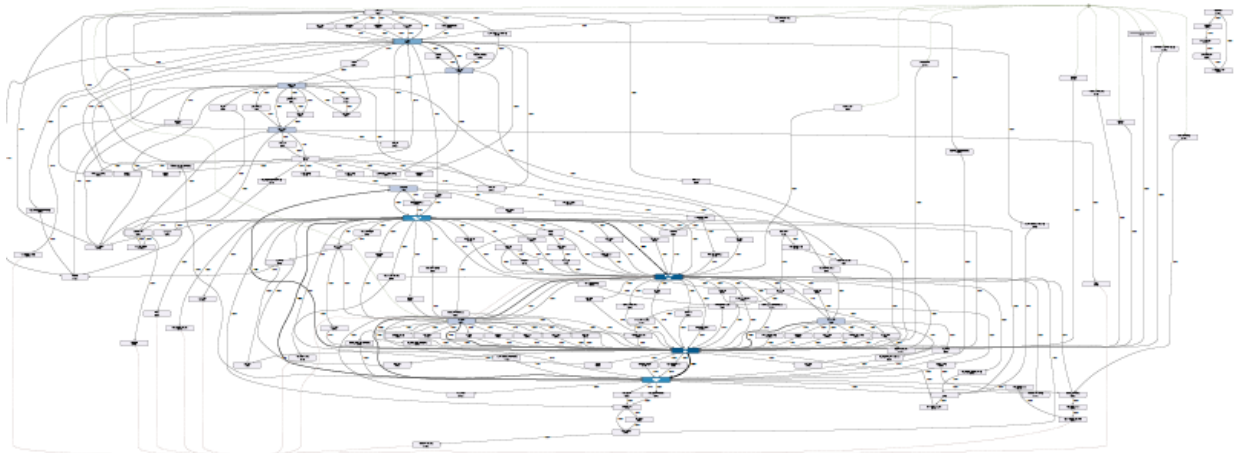

**Figure 10.** A snapshot of the partial process model discovered from the original event log.

## 7. Conclusion and Future Work

Existing process discovery algorithms assume that each activity label in the event log is unique and meaningful. The existence of redundant activity labels introduces unnecessary complexities to the discovered process models, which leads to "spaghetti-like" models. Hence, this paper proposes a multi-view framework to accurately detect redundant activity labels to produce more representative event logs for process mining. Our approach considers information from different views (i.e., control–flow relations, attribute values, and label semantic information) when detecting redundant activity labels. A consensus is made through the majority voting mechanism. The results are superior to those of the existing method in terms of detecting redundant activity labels. We also demonstrate the usability of our approach using a real-life EMR dataset.

In future work, first, we plan to investigate how different parameter settings can impact the method and to develop a method to automatically determine thresholds for different features. Second, we also aim to incorporate the NLP technique to automatically repair redundant activity labels by preserving the same contexts and by categorising differences according to their closest synonyms. Finally, we plan to investigate the feasibility of applying our approach in other domains.

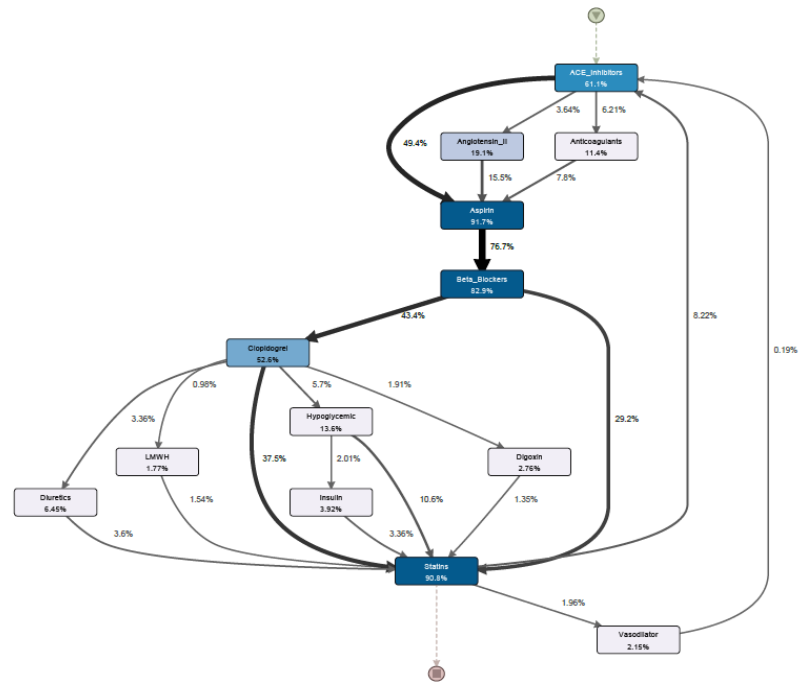

**Figure 11.** The process model discovered from the preprocessed event log.

**Author Contributions:** Conceptualisation and methodology, Q.C., Y.L., C.S.T., and S.K.P.; development, Q.C.; validation, Q.C., Y.L., C.S.T., and S.K.P.; writing—original draft preparation, Q.C.; writing—review and editing, Q.C., Y.L., C.S.T., and S.K.P.; supervision, C.S.T. and S.K.P. All authors have read and agreed to the published version of the manuscript.

**Funding:** This research received no external funding.

**Institutional Review Board Statement:** Ethics approval for the study (2019/ETH09692) was provided by the Northern Sydney Local Health District (NSLHD) Human Research Ethics Committee (HREC). Governance approvals were provided by the NSLHD and Central Coast Local Health District HRECs.

**Informed Consent Statement:** We received a waiver of consent for the patients in this study, which was approved by the NSLHD Human Research Ethics Committee. Only de-identified data were used by the researchers for this analysis, so a waiver of consent was appropriate.

**Data Availability Statement:** All datasets used in Section 5 to evaluate the proposed framework are publicly available. Please refer to notes for links to access the datasets. The datasets generated and/or analysed in Section 6 are not publicly available, as they are owned by the Chief Executives of the Local Health Districts and not by the researchers who performed this study.

**Acknowledgments:** The authors thank the members of the Speed-Extract research team and the technical assistance provided by the Sydney Informatics Hub, a Core Research Facility of the Univer-sity of Sydney. The authors also acknowledge ICT services at NSLHD. The authors also thank the Agency for Clinical Innovation, the NSW Ministry of Health, the Sydney Health Partners, and Health Innovation Funding for funding the Speed-Extract study.

**Conflicts of Interest:** The authors declare that there are no conflicts of interest.

## Notes